\documentclass{article} 
\usepackage{fullpage}
\usepackage{epsf}
\usepackage{amssymb}
 
\newtheorem{theorem}{Theorem}
\newtheorem{satz}[theorem]{Theorem}
\newtheorem{coro}[theorem]{Corollary}
\newcommand{\proof}{{\bf Proof:\ \ }}
\newtheorem{exam}[theorem]{Example}
\newtheorem{prop}[theorem]{Proposition}
\newtheorem{lemma}[theorem]{Lemma}

\newtheorem{definition}[theorem]{Definition}
\newcommand{\nat}{{\rm I \! N}}
\newcommand{\old}[1]{}
\newcommand{\clique}[1]{\mbox{$E_{#1}$}}

\newcommand{\qed}{\hfill\hbox{\quad $\Box$ \vspace{1ex}}}
\newcommand{\vol}{{\mbox{\it vol}}}
\def\part{\dot\cup}
\def\N{{\rm I\kern-.18em N}}
\def\Q{{\rm {\sf I}\kern-.42em Q}}
\def\Qp{\Q^{\scriptstyle +}}
\def\Qnn{\Qp_{\scriptstyle 0}}
\def\Z{{\mathchoice {\hbox{$\sf\textstyle Z\kern-0.4em Z$}}
    {\hbox{$\sf\textstyle Z\kern-0.4em Z$}}
    {\hbox{$\sf\scriptstyle Z\kern-0.3em Z$}}
    {\hbox{$\sf\scriptscriptstyle Z\kern-0.2em Z$}}}}

\begin{document}

\title{A General Framework for Bounds for Higher-Dimensional Orthogonal
Packing Problems\thanks{A previous extended abstract version of 
this paper appears in
{\em Algorithms -- ESA'97}\cite{esa}}}
 
\author{S\'andor P. Fekete\\
Department of Mathematical Optimization\\
Braunschweig University of Technology\\
D--38106 Braunschweig\\
GERMANY\\
        {\tt s.fekete@tu-bs.de}\and
        J\"org Schepers\thanks{
Supported by the German Federal Ministry of Education,
Science, Research and Technology (BMBF, F\"orderkennzeichen 01~IR~411~C7).}\\
        IBM Germany\\
        Gustav-Heinemann-Ufer 120/122\\
        D--50968 K\"oln\\
        GERMANY\\
        {\tt schepers@de.ibm.com}
}

\date{}
\maketitle 

\begin{abstract}

Higher-dimensional orthogonal packing problems have a wide range of 
practical applications, including packing, cutting, and scheduling.
In the context of a branch-and-bound framework for solving
these packing problems to optimality, it is of crucial importance
to have good and easy bounds for an optimal solution.
Previous efforts have produced a number of special classes
of such bounds. Unfortunately, some of these bounds are somewhat
complicated and hard to generalize. 
We present a new approach for obtaining classes of lower bounds
for higher-dimensional packing problems; our bounds improve
and simplify several well-known bounds from previous
literature. In addition, our approach provides an easy framework
for proving correctness of new bounds.

This is the second in a series of four articles describing
new approaches to higher-dimensional packing.
\end{abstract}

\section{Introduction}
\label{sec:intro}
The problem of cutting a rectangle into smaller rectangular 
pieces of given sizes is known as 
the {\em two--dimensional cutting stock problem}. It arises in many 
industries, where 
steel, glass, wood, or textile materials are cut, but it also occurs 
in less obvious contexts, such as
machine scheduling or optimizing the layout of advertisements in newspapers.
The three-dimensional problem is important for practical 
applications as container 
loading or scheduling with partitionable resources. It can be 
thought of as packing axis-aligned
boxes into a container, with a fixed orientation of boxes. 
We refer to the generalized problem  in $d \ge 2$ dimensions as 
the {\em $d$-dimensional orthogonal knapsack problem (OKP-$d$)}.
Being a generalization of the bin packing problem,
the OKP-$d$ is strongly ${\cal NP}$-complete. 
The vast majority of work done in this field refers to a restricted problem, 
where only
so--called  {\em guillotine patterns} are permitted.
This constraint arises from certain industrial cutting applications: 
guillotine patterns are those packings that can be generated by
applying a sequence of edge-to-edge cuts.
The recursive structure of these patterns makes
this variant much easier to solve than the {\em general}
 or {\em non--guillotine} problem. 

A common approach for obtaining bounds for geometric packing problems
arises from the total volume of the items -- if it exceeds the
volume of the container, the set cannot be packed.
Thus, we get a one-dimensional problem and do not have to 
consider the structure of possible packings.
These bounds are easy to achieve; however, they tend to be rather crude.
Just like in the one-dimensional case (see \cite{FESE97} for
an overview and a discussion), there have been attempts to improve these bounds.
In higher dimensions, this is somewhat more complicated; see
\cite{MAPV97,MAVI96} for two successful approaches.
However, higher-dimensional bounds are still hand-taylored,
somewhat complicated and hard to generalize. 

In this paper, we propose a generalization of this
method that leads to better results.
The basic idea is to use a number of volume tests, after
modifying the sizes of the boxes. The transformation
tries to reflect the relative ``bulkiness'' of the items,
and the way they can be combined.

This is the second in a series of four paper describing
new approaches to higher-dimensional orthogonal packing. \cite{pack1,fs-cchdop-04}
presents a combinatorial charachterization of feasible packings,
which is the basis for an effective branch-and-bound approach.
A preliminary version of the present paper was \cite{pack2}.
The third paper \cite{pack3,fs-eahdop-04} describes a resulting
overall algorithm. The more recent \cite{pack4} considers
higher-dimensional packing in the presence of order constraints.

The rest of this paper is organzied as follows.
After some basic definitions and notation in Section~\ref{sec:prelim},
Section~\ref{sec:dual} introduces the fundamental concept of
dual feasible functions. Section~\ref{sec:CS} describes how to use
these functions for the construction of conservative scales, which
yield a formal basis for lower bounds. Section~\ref{sec:BfOPP}
presents a number of lower bounds for various higher-dimensional
packing problems; Section~\ref{sec:BfPC} shows how to apply them to
{\em packing classes}, a concept that was introduced in 
\cite{pack1,fs-cchdop-04}.

\section{Preliminaries}
\label{sec:prelim}

\subsection{Basic Setup}
In the following, we consider a set of 
$d$-dimensional {\em boxes}
that need to be packed into a {\em container}. 

The input data is a finite set of boxes $V$, 
and a (vector-valued) {\em size function}
$w: V \rightarrow {\Qnn}^{d}$ that describes the size
of each box in any dimension $x_1,\ldots, x_n$.
For the orthogonal knapsack problem (OKP), we also have 
a value function $v: V \rightarrow {\Qp}$ that describes
the objective function value for each box.


The size of the container is given by a vector
$W \in {\Q^{+}}^{d}$. Whenever convenient,
we may assume that
the container is a $d$-dimensional unit cube.
Without loss of generality, we assume that each individual box fits
into the container, i.\,e., $ w(b) \le W $ holds for each box. 

For the volumes of boxes and container we use the following notation.
If $b \in V$, and $w$ is a size function defined on $V$, then
$
\vol_w(b) := \prod_{i=1}^{d} w_{i}(b)
$
denotes the volume of box $b$ with respect to $w$. Similarly, the
volume of the container is denoted by
$
\vol_W := \prod_{i=1}^{d} W_{i}.
$

If $S$ is a finite set and $f$ a real-valued function on $S$,
then we use the abbreviation
$
f(S) := \sum_{x \in S} f(x).
$

\subsection{Orthogonal Packings}
We consider arrangements of boxes that satisfy the following constraints:
\begin{enumerate}
\item {\bf Orthogonal Packing:} Each face of a box is parallel to
a face of the container.
\item {\bf Closedness:} No box may exceed the boundaries of the container.
\item {\bf Disjointness:} No two boxes must overlap. 
\item {\bf Fixed Orientations:} The boxes must not be rotated.
\end{enumerate}
In the following, we imply these conditions when ``packing boxes into 
a container'', ``considering a set of boxes that fits into a
container'', and speak of {\em packings}.

In the following, we assume that the position of any box is
given by the coordinate of the corner that is closest to the origin.

\subsection{Objective Functions}
Depending on the objective function, we distinguish 
three types of orthogonal packing problems:

\begin{itemize}
\item The {\bf Strip Packing Problem (SPP)} asks for the minimal height
$W_d$ of a container that can hold all boxes, where the size in the other
$d-1$ dimensions $W_1,\ldots, W_{d-1}$ are fixed.
\item 
For the 
{\bf Orthogonal Bin Packing Problem (OBPP)}, we have to determine
the minimal number of identical containers that are required
to pack all the boxes.
\item In the {\bf Orthogonal Knapsack Problem (OKP)},
each box has an objective value. A container has to be packed,
such that the total value of the packed boxes is maximized.
\end{itemize}

To clarify the dimension of a problem, we may speak of
OKP-$2$, OKP-$3$, OKP-$d$, etc.

Problems SPP and OBPP are closely related.
For $d \in \N$, an OBPP-$d$ instance can be transformed into a special
type of SPP-$(d+1)$ instance, by assigning the same $x_{d+1}$-size
to all boxes. This is of some importance for deriving relaxations
and lower bounds.

For all orthogonal packing problems we have to satisfy the constraint
that a given set of boxes fits into the container. This underlying 
decision problem is of crucial importance for our approach.

\begin{itemize}
\item {\bf Orthogonal Packing Problem (OPP)}:
Decide whether a set of boxes $V$ can be packed into the container.
\end{itemize}

\section{Dual Feasible Functions} 
\label{sec:dual}
The main objective of this paper is 
to describe good criteria for dismissing a candidate set 
of boxes. We will use the volume criterion on transformed volumes, 
by transforming volumes in a way
that any {\em transformed} instance can still be packed,
if the original instance could be packed. For this purpose,
we describe higher-dimensional transformations called
{\em conservative scales}. A particular way of getting
conservative scales is to construct them from (one-dimensional)
{\em dual feasible functions.}

For the rest of this paper, we assume without loss of generality
that the items have size $x_i\in [0,1]$, and the container size $W$ is
normalized to 1. Then we introduce the following:
 
\begin{definition}[Dual Feasible Functions]
A function $u : [0,1] \rightarrow [0,1]$ is called {\em dual feasible},
if for any finite set $S$ of nonnegative real numbers, we have the relation
\begin{equation} \label{dualzul}
\sum_{x \in S} x \leq 1 \, \Longrightarrow \,
\sum_{x \in S} u(x) \le 1.
\end{equation}
\end{definition}
Dual feasible functions have been
used in the performance analysis of heuristics for the BPP,
first by Johnson~\cite{JOHN73}, then by Lueker~\cite{lueker};
see Coffman and Lueker~\cite{COLU91} for a more detailed description.
The term (which was first introduced by Lueker~\cite{lueker})
refers to the fact that
for any dual feasible function $u$ and for any
bin packing instance with item sizes
$x_{1},\dots,x_{n}$, the vector
$(u(x_{1}),\dots,u(x_{n}))$ is a feasible solution
for the dual of the corresponding fractional bin packing problem
(see~\cite{KK82}).
By definition, convex combination and compositions of dual feasible
functions are dual feasible.
 
Dual feasible functions can be used
for improving lower bounds for the one-dimensional bin packing problem.
This is based on the following easy lemma.
 
\begin{lemma} \label{ubausdzf}
Let $I:=(x_{1},\dots,x_{n})$ be a BPP instance and let $u$ be
a dual feasible function.
Then any lower bound for the transformed BPP instance
$u(I):=(u(x_{1}),\dots,u(x_{n}))$ is also a lower bound for $I$.
\end{lemma}
 
By using a set of dual feasible functions
${\cal U}$ and considering the maximum value
over the transformed instances
$u(I), \, u \in {\cal U}$, we can try to obtain even better lower bounds.
 
In our paper \cite{FESE97}, we describe how dual feasible functions 
can be used to obtain good classes of lower bounds for bin 
packing problems. Several of these functions will be used
in a higher-dimensional context later on, 
so we summarize the most important
results for the benefit of the reader.
For the BPP, we mostly try to increase the sizes by a dual feasible
function, since this allows us to obtain a tighter bound by using
the volume criterion.
The hope is to find a $u^{(k)}$ for which as many items
as possible are in the ``win zones'' -- the subintervals of
$[0,1]$ for which the difference is positive.
Given this motivation, each dual feasible function is 
illustrated by a figure showing its win and loss zones.

\begin{figure}[htbp]
\begin{center}
\leavevmode
\epsfxsize=.6\textwidth
\epsffile{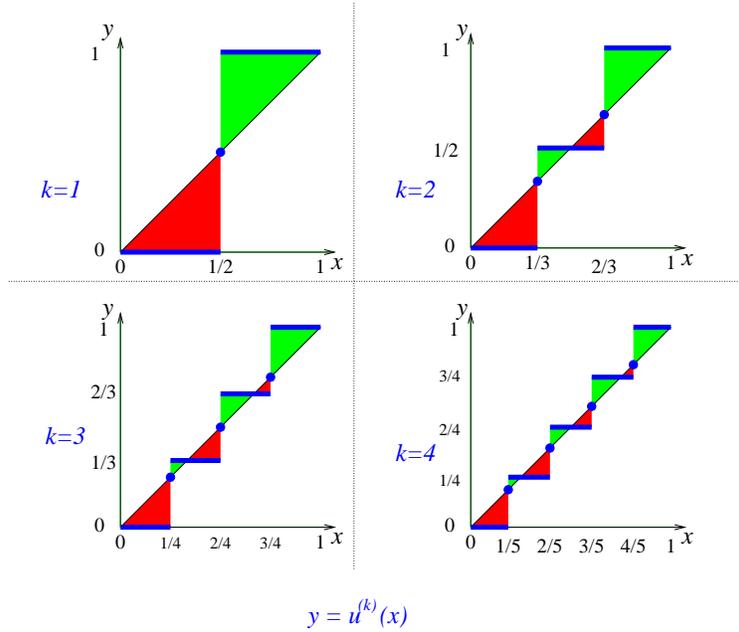}
\caption{Win and loss zones for $u^{(k)}$}
\label{dualzu}
\end{center}
\end{figure}

\begin{prop} \label{dfclass1}
Let $k \in \nat$. Then
\begin{eqnarray*}
u^{(k)}: [0,1]  & \rightarrow & [0,1] \\
x & \mapsto &
\left\{ \begin{array}{ll}
x, & \mbox{ for } x(k+1) \in \Z \\
\lfloor (k+1)x \rfloor \frac{1}{k}, & else
\end{array} \right.
\end{eqnarray*}
is a dual feasible function. (See Figure \ref{dualzu}.)
\end{prop}
 
\begin{figure}[htbp]
\begin{center}
\leavevmode
\epsfxsize=.3\textwidth
\epsffile{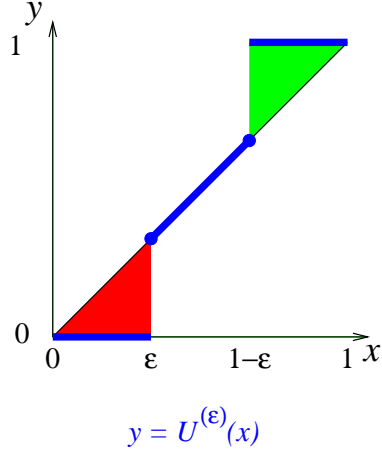}
\caption{Win and loss zones for $U^{(\epsilon)}$}
\label{dualzu2}
\end{center}
\end{figure}

The following class of dual feasible functions is the implicit basis
for the bin packing bound $L_2$ by
Martello and Toth~\cite{MATO90,MATO90b}.
This bound is obtained by neglecting all items smaller than
a given value $\epsilon$. We account for these savings by increasing
all items of size larger than $1-\epsilon$.
Figure~\ref{dualzu2} shows the corresponding win and loss zones.
 
\begin{prop} \label{dfclass2}
Let $\epsilon \in [0,\frac{1}{2}]$. Then
\begin{eqnarray*}
U^{(\epsilon)}: [0,1]  & \rightarrow & [0,1] \\
x & \mapsto & \left\{
\begin{array}{ll}
1, & \mbox{ for }x > 1-\epsilon \\
x, & \mbox{ for }\epsilon \leq x \leq 1-\epsilon  \\
0, & \mbox{ for }x < \epsilon
\end{array} \right.
\end{eqnarray*}
is a dual feasible function. (See Figure \ref{dualzu2}.)
\end{prop}
 
In our paper~\cite{FESE97}, we show how these two 
classes of dual feasible functions
can be combined by virtue of Lemma~\ref{ubausdzf}
in order to get good bounds for the one-dimensional BPP.
Our family of lower bounds can be computed in
time $O(n)$ after sorting the items by size,
and it dominates the class $L_{2}$ that was suggested by
Martello an Toth~\cite{MATO90,MATO90b} as a generalization
of the volume bound $L_1$.
Our framework of dual feasible functions
allows an easy generalization and proof of these bounds.
In \cite{FESE97},
we show that this generalized bound improves the asymptotic worst-case
performance from
$\frac{2}{3}$ to $\frac{3}{4}$, and provide empirical evidence
that also the practical performance is improved significantly.

\medskip
Our third class of dual feasible functions has some similarities
to some bounds that were hand-tailored for
the two-dimensional and three-dimensional
BPP by Martello and Vigo~\cite{MAVI96},
and Martello, Pisinger, and Vigo~\cite{MAPV97}. However,
our bounds are simpler and dominate theirs. (We will discuss
this in detail in Section~\ref{sec:BfOPP}.)
 
\begin{figure}[htbp]
\begin{center}
\leavevmode
\epsfxsize=.59\textwidth
\epsffile{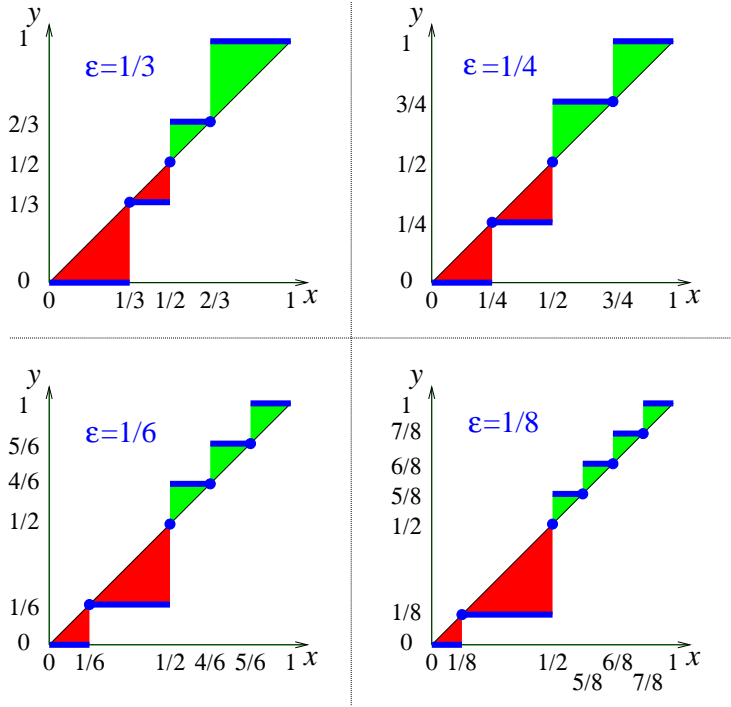}
\caption{Win and loss zones for $\phi^{(\epsilon)}$}
\label{dualzu3}
\end{center}
\end{figure}

This third class also ignores items of size below a threshold value
$\epsilon$. For the interval
$(\epsilon,\frac{1}{2}]$, these functions are constant,
on  $(\frac{1}{2},1]$ they have the form of step functions.
Figure~\ref{dualzu3} shows that for small values of
$\epsilon$, the area of loss zones for $\phi^{(\epsilon)}$
exceeds the area of win zones
by a clear margin.
This contrasts to the behavior of
the functions $u^{(k)}$ and
$U^{(\epsilon)}$, where the win and loss areas have the same size.
 
\begin{theorem} \label{dfclass3}
Let $\epsilon \in [0,\frac{1}{2})$. Then
\begin{eqnarray*}
\phi^{(\epsilon)}: [0,1]  & \rightarrow & [0,1] \\
x & \mapsto & \left\{
\begin{array}{ll}
1 - \frac{ \lfloor (1-x)\epsilon^{-1} \rfloor }{ \lfloor \epsilon^{-1} \rfloor}
 , & \mbox{ for }x > \frac{1}{2} \\
\frac{1}{\lfloor \epsilon^{-1} \rfloor}, & \mbox{ for }\epsilon \leq x \leq \frac{1}{2}  \\
0, & \mbox{ for }x < \epsilon
\end{array} \right.
\end{eqnarray*}
is a dual feasible function. (See Figure \ref{dualzu3}.)
\end{theorem}
 
\begin{proof}
Let $S$
be a finite set of nonnegative real numbers, with
$\sum\limits_{x \in S}x \leq 1$.
Let $S':=\{x \in S\,|\, \epsilon \leq x \leq \frac{1}{2} \}.$
We distinguish two cases:

If all elements of $S$ have size at most $\frac{1}{2}$, then by definition
of $S'$,
\begin{equation}
1 \ge
\sum_{x \in S} x
\ge 
\sum_{x \in S'} x
\ge
|S'| \epsilon
\end{equation}
holds. Since $|S'|$ is integral, it follows that
$|S'| \leq \lfloor \epsilon^{-1} \rfloor$, hence
\begin{equation}
\sum_{x \in S} \phi^{(\epsilon)}(x)
= 
\sum_{x \in S'} \phi^{(\epsilon)}(x)
=
|S'| \frac{1}{\lfloor \epsilon^{-1} \rfloor}
\le
1.
\end{equation}
Otherwise
$S$ contains exactly one element $y>\frac{1}{2}$ and we have
\begin{equation}
1 \ge
\sum_{x \in S} x
\ge 
y + \sum_{x \in S'} x
\ge
y + |S'| \epsilon.
\end{equation}
 
Therefore $|S'| \leq \lfloor (1-y) \epsilon^{-1} \rfloor$ and hence
\begin{equation}
\sum_{x \in S} \phi^{(\epsilon)}(x)
= 
\phi^{(\epsilon)}(y) +
\sum_{x \in S'} \phi^{(\epsilon)}(x)
=
1 - \frac{ \lfloor (1-y) \epsilon^{-1} \rfloor }{ \lfloor \epsilon^{-1} \rfloor}
 +
|S'| \frac{1}{\lfloor \epsilon^{-1} \rfloor}
\leq 1.
\end{equation}
\qed
\end{proof}

\section{Conservative Scales} 
\label{sec:CS}
In this section we return to our original goal:
deducing necessary conditions for feasible packings.

There can only be a packing if the total volume of the boxes does not exceed
the volume of the container. This trivial necessary criterion
is called the {\em volume criterion}. As we will see, it remains valid
even if we apply appropriate transformations on the box sizes.

For any coordinate $i$, let
$$
{\cal F}(V,w_{i}) := \{ S \subseteq V | w_{i}(S) \le 1 \},
 \, i \in \{1,\dots,d\}
$$
be the family of $i$-{\em feasible box sets}, i.e., the subsets of boxes
whose total $i$-width does not exceed the width of the container.
If we replace 
$w$ by a size function $w'$ without reducing these
families, then all packing classes remain intact:
\begin{satz} \label{bilanz}
Let $(V,w)$ and $(V,w')$ be OPP instances.
If for all
$i \in \{1,\dots,d\}$ we have
\begin{equation} \label{bilcond} 
{\cal F}(V,w_{i}) \subseteq
{\cal F}(V,w'_{i}),
\end{equation}
then any packing class for $(V,w)$ is also a packing class for $(V,w')$.
\end{satz}

\proof
As we showed in our paper~\cite{pack1}, the existence of a packing 
is equivalent to the existence of a packing class,
i.e., a set of $d$ graphs $G_i=(V, E_i), i=1,\ldots,d$
that have the following properties:
\begin{eqnarray*}
(P1):&& \mbox{Each $G_{i}:=(V,E_{i})$ is an interval graph.}
\\
(P2):&& \mbox{Each stable set } S \mbox{ of } G_{i} \mbox{ is $x_i$--feasible}.
\\
(P3):&& \bigcap_{i=1}^{d} E_{i} = \emptyset.
\end{eqnarray*}
(Intuitively, these graphs describe the overlap of the $i$-projections
of the boxes.)

This means we can focus on packing classes. 
The only condition on a packing class that
involves the size of the objects deals with condition (P2):
Any stable set of one of the component graphs $G_i$ should be
$i$-feasible, i.e., the total sum of $i$-widths should not exceed
the $i$-width of the container. This means that an OPP instance is
characterized by the families of $i$-feasible box sets.
By assumption, these are not changed when replacing $w$ by $w'$.
\qed

\begin{definition}[Conservative Scales]\label{konska}
A function $w'$ satisfying the conditions of Theorem~\ref{bilanz} 
is called a {\em conservative scale} for $(V,w)$.
\end{definition}
The desired generalization of the volume criterion follows
directly from Theorem~\ref{bilanz}:  
\begin{coro} \label{notkor}
If $w'$ s a conservative scale for the OPP-instance $(V,w)$, then
\begin{equation} \label{volumencond}
\sum_{b \in V} \vol_{w'}(b) \le 1
\end{equation}
is a necessary condition for the existence of a packing class 
for $(V,w)$.
\end{coro}
In order to apply the new criterion, we need a method for
constructing functions $w'$. A particular way of getting conservative scales
is given by the dual feasible functions described in the previous section.
\begin{lemma} \label{konskabydual}
Let $(V,w)$ be an OPP instance and
let $u_{1}, \dots, u_{d}$ be dual feasible functions.
Then $w' := (u_{1}(w_{1})$, $\dots$, $u_{d}(w_{d}))$ is
a conservative scale for $(V,w)$.
\end{lemma}
{\bf Proof:} 
For $i \in \{1,\dots,d\}$ the functions $u_{i}$ are dual feasible, hence
$$
S \in {\cal F}(V,w_{i}) 
\, \Leftrightarrow
\, \sum_{b \in S} w_{i}(b) \le 1
\, \Rightarrow
\, \sum_{b \in S} u_{i}(w_{i}(b)) \le 1
\, \Leftrightarrow
\, S \in {\cal F}(V,w'_{i}).
$$
\qed

\bigskip
In the previous section, we described the dual feasible functions
$u^{(k)}, \, k \in \N,$ that can be computed in constant time
for any item. They can be used to construct a conservative scale
for an OPP instance in $d \cdot |V|$-linear time. By checking the volume
criterion for a small set of conservative scales, we get a fast heuristic
method for identifying OPP instances without a feasible packing.

\begin{exam} \label{neunmonkeys}  Consider the three-dimensional OPP
\vskip12pt

``Do nine cubes of size $\frac{2}{5}$ fit into a unit cube container?''.
\vskip12pt

The total volume is $9 \times \frac{8}{125} = 0.576 \le 1$, hence
the volume criterion does not produce an answer.
By applying the dual feasible function $u^{(2)}$ to all components, 
we get a conservative scale. The transformed boxes are cubes of size 
$\frac{1}{2}$.
Now the total volume is $9 \times \frac{1}{8} = 1.125 > 1$, so 
with the help of Corollary~\ref{notkor}, we get the answer
``no'' to the original question.
\end{exam}
\begin{figure}[htbp]
\begin{center}
\leavevmode
\epsfxsize=9cm
\epsffile{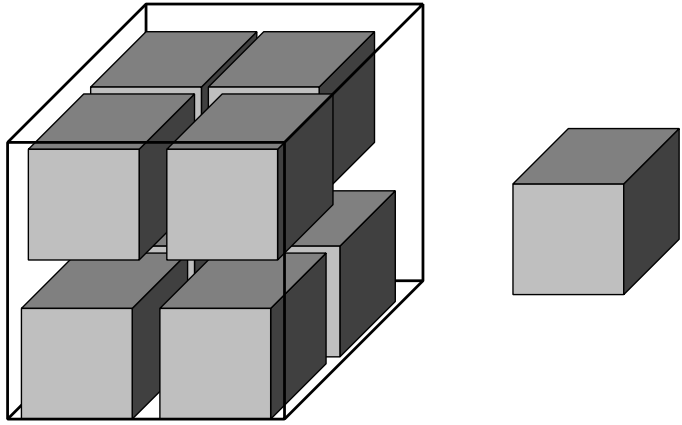}
\caption{Trying to fit nine cubes of size $0.4$ into a unit container. (Example~\ref{neunmonkeys})}
\label{neuenmonkeysabb}
\end{center}
\end{figure}

\section{Bounds for Orthogonal Packing Problems} 
\label{sec:BfOPP}
Corollary \ref{notkor} 
yields a generic method for deriving linear relaxations
of orthogonal packing problems:
Let ${\cal W}$ be an arbitrary set of conservative scales for
$(V,w)$. Then the difficult constraint
$$
\mbox{``There is a packing class for }(S,w)\mbox{''}
$$
can be replaced by the linear restriction
$$
\forall w' \in {\cal W}: \, \vol_{w'}(S) \le 1.
$$
The corollary implies that the latter is a relaxation.
In the following we will use this method to obtain bounds
for all orthogonal packing problems.

\subsection{Strip Packing Problem} \label{spprelaxsec}
For the SPP-$d$, we get a lower bound:
\begin{lemma}\label{lbspp}
Let $V$ be a set of boxes, let $w$ be a size function on $V$,
and let ${\cal W}$ be a finite set of conservative scales for $(V,w)$. Then
the function $L_{SPP}(V,w)$, defined by
$$
L_{SPP}(V,w) := \max_{w' \in {\cal W}} \vol_{w'}(V),
$$
is a lower bound for the SPP instance $(V,w)$.
\end{lemma}
The proof is immediate.

If a conservative scale $w'$ is composed from 
dual feasible functions as described in Section~\ref{sec:dual},
which can be evaluated in constant time for each box,
then $L_{SPP}(V,w')$ can be computed in time that is linear in
$d \cdot |V|$.

\subsection{Orthogonal Knapsack Problem} \label{okprelaxsec}
By using the above relaxation, the OKP turns into a
{\em higher-dimensional knapsack problem} (MKP) 
(see \cite{FRPL94}).
While the OKP is strongly {\cal NP}-hard,
the MKP can be solved in pseudopolynomial time by
using the dynamic program for the one-dimensional
problem.

In case of ${\cal W} = \{w'\}$ we get a one-dimensional
knapsack problem as an OKP relaxation, with
item weights given by box volumes with respect to the
conservative scale $w'$, and capacity given by the volume
of the container. These bounds are used in the implementation
of our OKP algorithm, which is described in detail
in our paper \cite{pack3}

\subsection{Orthogonal Bin Packing Problem} \label{obpplb}
The OBPP is relaxed to a
{\em Vector Packing Problem} (see \cite{GAKW93},\cite{GGJY76}).
This problem is still strongly {\cal NP}-hard.

In order to derive a bound for the OBPP that is easier to compute,
we consider the continuous lower bound
$L_{0}$ for the one-dimensional BPP. For unit capacity, this bound arises as the 
sum of item sizes, rounded up to the next integer.
$L_0$ yields a good approximation of the optimum, as long
as there are sufficiently many ``small'' items 
(see \cite{MATO90,MATO90b}).

For many instances, a majority of boxes has at least one dimension
that is significantly smaller than the size of the container.
This means that the box volume is significantly smaller than the
volume of the container.
In general, this is still true after a transformation by a conservative scale.
This makes it plausible to use $L_0$ on transformed volumes.
Summarizing:
\begin{lemma}\label{lbooo}
Let $V$ be a set of boxes, let $w$ be a size function on $V$,
and let ${\cal W}$ be a finite set of
conservative scales for $(V,w)$. Then
the function $L_{OBPP}(V,w)$, defined by
$$
L_{OBPP}(V,w) := \max_{w' \in {\cal W}} \lceil \vol_{w'}(V) \rceil,
$$
is a lower bound for the normalized OBPP instance $(V,w)$.
\end{lemma}

For 
conservative scales $w'$ that are composed from dual feasible functions
as described in Section \ref{sec:dual},
$L_{OBPP}(V,w')$ can be computed in time linear in
$d \cdot |V|$ --- just 
like the  bounds for strip packing from Lemma \ref{lbspp}.

We demonstrate the advantages of our method
by showing that generalizations of the lower bounds for the OBPP-$2$ and the 
OBPP-$3$ from the papers by Martello and Vigo \cite{MAVI96} and
Martello, Pisinger, and Vigo \cite{MAPV97} fit into this framework.
Those bounds are stated as maxima of several partial bounds
that can all be stated as in Lemma \ref{lbooo}. 
The necessary conservative scales can all be constructed from dual feasible
functions $u^{(1)},U^{(\epsilon)}$ and $\phi^{(\epsilon)}$.
In \cite{MAVI96} and \cite{MAPV97}, these partial bounds are all derived
by separate considerations, some of which are quite involved.
We will discuss two of these cases to demonstrate how our framework
leads to a better understanding of bounds and helps to find improvements.
Particularly useful for this purpose is the decomposition
of bounds into independent one-dimensional components,
which was established in Lemma \ref{konskabydual}.

We state the original formulation of the most complicated partial bound in \cite{MAVI96}.
The only modification (apart from the naming of variables) arises from applying their
bound to the normalized OBPP-$2$. (The integrality of input data that is
required in \cite{MAVI96} is of no consequence in this context.)
For two parameters $p,q \in (0,\frac{1}{2}]$, Martello and Vigo set
\begin{eqnarray*}
I_{1}(p,q) & := & \left\{ b \in V \quad | \quad w_{1}(b) > 1-p \quad \wedge \quad w_{2}(b) > 1-q \right\}, \\
I_{2}(p,q) & := & \left\{ b \in V \setminus I_{1}(p,q) \quad | \quad w_{1}(b) > \frac{1}{2} \quad \wedge \quad w_{2}(b) > \frac{1}{2} \right\},\\
I_{3}(p,q) & := & \left\{ b \in V \quad | \quad \frac{1}{2} \ge w_{1}(b) \ge p \quad \wedge \quad \frac{1}{2} \ge w_{2}(b) \ge q  \right\},
\end{eqnarray*}
$$
m(b,p,q) := 
\left\lfloor \frac{1}{p} \right\rfloor \left\lfloor \frac{1-w_{2}(b)}{q} \right\rfloor
+ \left\lfloor \frac{1}{q} \right\rfloor \left\lfloor \frac{1-w_{1}(b)}{p} \right\rfloor
- \left\lfloor \frac{1-w_{1}(b)}{p} \right\rfloor \left\lfloor \frac{1-w_{2}(b)}{q} \right\rfloor,
$$
$$
L^{(p,q)}(V,w) := | I_{1}(p,q) \cup I_{2}(p,q) | +  
\left\lceil \frac{|I_{3}(p,q)| - \sum_{b \in I_{2}(p,q)} m(b,p,q) }{ 
\lfloor p^{-1} \rfloor  \lfloor q^{-1} \rfloor} \right\rceil. 
$$
Martello and Vigo derive the bound $L^{(p,q)}$  
(which is denoted by $L_{3}(p,q)$ in \cite{MAVI96})   
by a geometric argument that uses a certain type of normal form of two-dimensional
packings. This implies that only boxes from the sets
$I_{1}(p,q)$, 
$I_{2}(p,q)$, $I_{3}(p,q)$ are taken into account.

We now give a formulation according to Lemma \ref{lbooo}
that uses a conservative scale.
\begin{satz} \label{vigosatz2d}
Let $p,q \in (0,\frac{1}{2}]$. With 
$$
w'^{(p,q)} := (\phi^{(p)} \circ w_{1}, \phi^{(q)} \circ w_{2})
$$
we have
$$
L^{(p,q)}(V,w) = \lceil \vol_{w'^{(p,q)}}(I_{1}(p,q) \cup I_{2}(p,q) \cup I_{3}(p,q)) \rceil.
$$
\end{satz}
{\bf Proof.}

In the following we only use elementary transformations and the definition
of the dual feasible functions $\phi^{(\epsilon)}$ from Theorem~\ref{dfclass3}.

By definition of
$I_{1}(p,q)$ and 
$I_{3}(p,q)$ it follows immediately that
$$\forall b \in I_{1}(p,q): \vol_{w'^{(p,q)}}(b) = 1 \cdot 1 = 1$$
and
$$\forall b \in I_{3}(p,q):
\vol_{w'^{(p,q)}}(b) = \frac{1}{\lfloor p^{-1} \rfloor} \frac{1}{\lfloor q^{-1} \rfloor}.
$$
Furthermore, we get for $b \in I_{2}(p,q)$ 
\begin{eqnarray*}
&& \vol_{w'^{(p,q)}}(b) \quad    
 = \quad  \phi^{(p)}(w_{1}(b)) \cdot
\phi^{(q)}(w_{2}(b)) \\
& = &
\left(
1 - \frac{\lfloor (1-w_{1}(b)) p^{-1}\rfloor}
{ \lfloor p^{-1} \rfloor}
\right)
\left(
1 - \frac{\lfloor (1-w_{2}(b)) q^{-1}\rfloor}{ \lfloor q^{-1} \rfloor}
\right) \\
& = &
1 - \frac{\lfloor (1-w_{1}(b)) p^{-1}\rfloor}
{ \lfloor p^{-1} \rfloor}
  - \frac{\lfloor (1-w_{2}(b)) q^{-1}\rfloor}
{ \lfloor q^{-1} \rfloor}
+ \frac{\lfloor (1-w_{1}(b)) p^{-1}\rfloor
      \lfloor (1-w_{2}(b)) q^{-1}\rfloor}{\lfloor p^{-1} \rfloor \lfloor q^{-1} \rfloor}\\
& = &
1 - \frac{
\left(
\left\lfloor \frac{1}{q} \right\rfloor \left\lfloor \frac{(1-w_{1}(b))}{p}\right\rfloor +
\left\lfloor \frac{1}{p} \right\rfloor \left\lfloor \frac{(1-w_{2}(b))}{q}\right\rfloor -
\left\lfloor \frac{(1-w_{1}(b))}{p}\right\rfloor
\left\lfloor \frac{(1-w_{2}(b))}{q}\right\rfloor
\right)}
{\lfloor p^{-1} \rfloor \lfloor q^{-1} \rfloor} \\
& = &
1 - \frac{m(b,p,q)}{\lfloor p^{-1} \rfloor \lfloor q^{-1} \rfloor}. 
\end{eqnarray*}
All in all, this yields:
\begin{eqnarray*}
&& \lceil \vol_{w'^{(p,q)}}(I_{1}(p,q) \cup I_{2}(p,q) \cup I_{3}(p,q)) \rceil \\
& = &
\left\lceil 
\sum_{b \in I_{1}(p,q)} \vol_{w'^{(p,q)}}(b) 
+ \sum_{b \in I_{2}(p,q)} \vol_{w'^{(p,q)}}(b) 
+ \sum_{b \in I_{3}(p,q)} \vol_{w'^{(p,q)}}(b) 
\right\rceil 
\\
& = &
\left\lceil 
\sum_{b \in I_{1}(p,q)} 1 
+ \sum_{b \in I_{2}(p,q)}  
\left(1 - \frac{m(b,p,q)}{\lfloor p^{-1} \rfloor \lfloor q^{-1} \rfloor} \right)
+ \sum_{b \in I_{3}(p,q)}  
\frac{1}{\lfloor p^{-1} \rfloor \lfloor q^{-1} \rfloor}
\right\rceil 
\\
& = &
\left\lceil 
| I_{1}(p,q) | +
| I_{2}(p,q) | 
- \sum_{b \in I_{2}(p,q)}  
\frac{m(b,p,q)}{\lfloor p^{-1} \rfloor \lfloor q^{-1} \rfloor}
+ \frac{I_{3}(p,q)}{\lfloor p^{-1} \rfloor \lfloor q^{-1} \rfloor}
\right\rceil 
\\
& = &
L^{(p,q)}(V,w).
\end{eqnarray*}
\qed

\bigskip
This alternative formulation of
$L^{(p,q)}$ from Theorem \ref{vigosatz2d}
reveals a significant disadvantage of the bound:
Boxes that are not contained in one of the sets
$I_{1}(p,q), I_{2}(p,q), I_{3}(p,q)$ are disregarded for the balance,
even if their volume is positive with respect to $w'^{(p,q)}$.
In particular, boxes are disregarded that are narrow in one
coordinate direction 
($w_{i}(b) \le \frac{1}{2}$) and wide in the other
($w_{i}(b) > \frac{1}{2}$).
For example, the OBPP-$2$ instance with
$$
V:=\{b_{1},b_{2},b_{3}\}, \quad w(b_{1}):=\left(\frac{2}{3},\frac{1}{2}\right), \,
w(b_{2})=w(b_{3}):=\left(\frac{1}{2},\frac{2}{3}\right)
$$
yields
$L^{(p,q)}(V,w) = 0$
for all
$p,q \in (0,\frac{1}{2}]$. None of the other partial bounds
from \cite{MAVI96} exceed the value of $1$.
However, the above discussion yields the following improvement of 
$L^{(p,q)}$:
$$
L'^{(p,q)}(V,w) = \lceil \vol_{w'^{(p,q)}}(V) \rceil
$$
This yields $L'^{(\frac{1}{2},\frac{1}{2})}(V,w) = 2$, which is the optimal value.

We now give a description of the OBPP bound from \cite{MAVI96} with the above
improvement in the framework of conservative scales.
Like the proof of 
Theorem~\ref{vigosatz2d}, it can be derived by using only 
the definition of conservative scales and elementary transformations.
\begin{theorem}
For $p, q \in (0,\frac{1}{2}]$ let
$$                
\begin{array}{lccrrc}
w^{(1)(p)} & := &(& u^{(1)} \circ w_{1},&  U^{(p)} \circ w_{2}&), \\
w^{(2)(p)}  & := &(& U^{(p)} \circ w_{1},&  u^{(1)} \circ w_{2}&), \\
[1ex]
w^{(3)(p)} & := &(& u^{(1)} \circ w_{1},&  \phi^{(p)} \circ w_{2}&), \\
w^{(4)(p)}  & := &(& \phi^{(p)} \circ w_{1},&  u^{(1)} \circ w_{2}&), \\
[1ex]
w^{(5)(p)}  & := &(& w_{1},&  U^{(p)} \circ w_{2}&),\\
w^{(6)(p)}  & := &(& U^{(p)} \circ w_{1},&  w_{2}&),\\
[1ex]
w^{(7)(p,q)}  & := &(& \phi^{(p)} \circ w_{1},&  \phi^{(q)} \circ w_{2}&).
\end{array}
$$                        
By Lemma \ref{lbooo}, the maximum over the partial bounds is
$$
L_{2d} :=  
\max \left\{
\max_{k \in \{1,\dots,6\}}
\max_{0<p \le \frac{1}{2}} 
\{ \lceil \vol_{w^{(k)(p)}}(V) \rceil \},
\max_{0<p \le \frac{1}{2}, \, 0<q \le \frac{1}{2}} \{ \lceil \vol_{w^{(7)(p,q)}}(V) \rceil \}
\right\}.
$$
$L_{2d}$ dominates the bound $L_{4}$ from \cite{MAVI96}.
\end{theorem}
Like for the bound $L_2$ (see Martello and Toth \cite{MATO90}),
finding the maxima over the parameters $p,q$ can be 
reduced to considering finitely many values. This makes it 
possible to compute the one-parametric bounds in time
$O(|V|)$, and the two-parametric bound in time $O(|V|^{2})$,
provided that the boxes are sorted by size for each coordinate direction.

Most partial bounds for the OKP-$3$ that are given in
\cite{MAPV97} disregard boxes that do not span at least half the container
volume in two coordinate directions. Since these boxes have to be stacked in the
third direction, it is possible to use the corresponding bounds for the
one-dimensional BPP.

Not counting symmetry, there is only one partial bound
in \cite{MAPV97} that is really based on higher-dimensional
properties. For this bound, we derive a dominating new bound in terms of conservative
scales.

For this purpose, let $p,q \in (0,\frac{1}{2}]$ and consider the sets
\begin{eqnarray*}
J_{1}(p,q) & := & \left\{ b \in V \quad | \quad w_{1}(b) > 1-p \quad \wedge \quad w_{2}(b) > 1-q \right\}, \\
J_{2}(p,q) & := & \left\{ b \in V \setminus J_{1}(p,q) \quad | \quad w_{1}(b) > p \quad \wedge \quad w_{2}(b) > q \right\}.
\end{eqnarray*}
A simplified formulation of the lower bound for the normalized
OBPP-$3$ is
$$
L^{(p,q)}(V,w) :=   
\left\lceil  
\sum_{b \in J_{1}(p,q)} w_{3}(b) + \sum_{b \in J_{2}(p,q)} \vol_{w}(b)
\right\rceil.  
$$
In the first two coordinate directions, the size of the box is rounded to the full
size of the container, if and only if {\em both}
sizes exceed their respective threshold values
$(1-p)$ and $(1-q)$. Noting that the same effect can be reached
in each dimension by using the dual feasible function
$U^{(\epsilon)}$, we see that the interconnection between 
both directions is unnecessary.
Using the conservative scale
$$
w'^{(p,q)} := ( U^{(p)} \circ w_{1}, U^{(q)} \circ w_{2}, w_{3}),
$$
we can round up both sizes in the same way, 
but independent from each other.
This yields the following bound that dominates $L^{(p,q)}$: 
$$
L'^{(p,q)}(V,w) :=   
\left\lceil  
w'^{(p,q)}(V)
\right\rceil. 
$$
For an example that there is some improvement,
consider the 
OBPP-$3$ instance $(V,w)$ with five boxes
of sizes
$\left(\frac{2}{3},\frac{1}{2},\frac{1}{2}\right)$.
For all
$p,q \in (0,\frac{1}{2}]$ 
the set $J_{1}(p,q)$ is empty, so that
$L^{(p,q)}(V,w) \le \lceil \vol_{w}(V) \rceil = \left\lceil \frac{5}{6} \right\rceil = 1$ holds. 
On the other hand,
$L'^{(p,q)}(V,w) = \lceil \frac{5}{4} \rceil = 2$ yields the optimal value. 

All other partial bounds from
\cite{MAPV97} can be formulated as in Lemma \ref{lbooo}.
The proof is left to the reader.
\begin{theorem}
Using the following conservative scales, we get a description of the bound
from \cite{MAPV97}, including the improvement mentioned above.
$$
\begin{array}{lccrrrc}
w^{(1)(p)} & := &(& u^{(1)} \circ w_{1},&  u^{(1)} \circ w_{2},& U^{(p)} \circ w_{3}&) ,\\
w^{(2)(p)} & := &(& u^{(1)} \circ w_{1},&  U^{(p)} \circ w_{2}, & u^{(1)} \circ w_{3}&),\\
w^{(3)(p)} & := &(& U^{(p)} \circ w_{1},& u^{(1)} \circ w_{2},&  u^{(1)} \circ w_{3}&),\\
[1ex]
w^{(4)(p)} & := &(& u^{(1)} \circ w_{1},&  u^{(1)} \circ w_{2},& \phi^{(p)} \circ w_{3}&),\\
w^{(5)(p)} & := &(& u^{(1)} \circ w_{1},&  \phi^{(p)} \circ w_{2}, & u^{(1)} \circ w_{3}&),\\
w^{(6)(p)} & := &(& \phi^{(p)} \circ w_{1},& u^{(1)} \circ w_{2},&  u^{(1)} \circ w_{3}&),\\
[1ex]
w^{(7)(p,q)}  & := &(& U^{(p)} \circ w_{1},&  U^{(q)} \circ w_{2},&  w_{3}&),\\
w^{(8)(p,q)}  & := &(& U^{(p)} \circ w_{1},&  w_{2},& U^{(q)} \circ w_{3}&),\\
w^{(9)(p,q)}  & := &(& w_{1},& U^{(p)} \circ w_{2},& U^{(q)} \circ w_{3}&).     
\end{array}
$$
Then the maximum of the corresponding partial bounds according to Lemma \ref{lbooo} is
$$
L_{3d}(V,w) :=  
\max \left\{ 
\max_{k \in \{1,\dots,6\}}
\max_{0<p \le \frac{1}{2}} 
\{ \lceil \vol_{w^{(k)(p)}}(V) \rceil \},
\max_{k \in \{7,\dots,9\}}
\max_{0<p \le \frac{1}{2}, \, 0<q \le \frac{1}{2}} \{ \lceil \vol_{w^{(k)(p,q)}}(V) \rceil \}
.\right\}
$$
This dominates the bound $L_{2}$ from \cite{MAPV97}.
\end{theorem}
Like $L_{2d}$, $L_{3d}$ can be computed in quadratic time.

From the new formulation of $L_{2d}$ and $L_{3d}$ it is immediate
how the bounds can be improved by applying further dual feasible functions,
without changing the complexity.
$u^{(1)},U^{(\epsilon)}$, and $\phi^{(\epsilon)}$ are all dual feasible functions
with win zones in the interval 
$(\frac{1}{2},1]$. (See Figures \ref{dualzu}, \ref{dualzu2}, \ref{dualzu3}). 
This suggests it may be a good idea to use 
$u^{(k)}, k \ge 2$ as well, where 
sizes $\le \frac{1}{2}$ may be rounded up.

\section{Bounds for Packing Classes}
\label{sec:BfPC}
As mentioned in the proof of Theorem~\ref{bilanz}, and described
in detail in our paper~\cite{pack1}, it suffices to construct
packing classes instead of explicit feasible geometric arrangements
of boxes. This allows us to enumerate over feasible packings
by performing a branch-and-bound scheme over the edge sets
of the component graphs $G_i=(V,E_i)$.
At each stage of this scheme, we have fixed a subset of edges
${\cal E}_{+,i} \subseteq E_{i}, \, i=1,\dots,d$
to be in the $i$th component graph $G_i$.

In the following, we describe how conservative scales can be used
for constructing bounds on partial edge sets.
In our exact algorithm that is described in \cite{pack3},
they help to limit the growth of the search tree.

\bigskip
First of all, conservative scales can be used to weaken
the assumptions of Theorem~9 from our paper \cite{pack1}: 
\begin{theorem} \label{pmclique2}
Let $E$ be a packing class for $(V,w)$, 
$w'$ be a conservative scale for $(V,w)$,
$i \in \{1,\dots,d\}$
and
$S \subseteq V$. Let $G_{i}$ be the $i$th component graph of $E$.
Then $G_{i}[S]$ contains a clique of cardinality
$\lceil w'_{i}(S)\rceil$.
\end{theorem} 
{\bf Proof:} 
By Theorem~\ref{bilanz}, any packing class 
$E$ for $(V,w)$ is also a packing class for
$(V,w')$. Then the claim follows from Theorem~9 
in \cite{pack1}.
\qed

\bigskip
Now we generalize conservative scales.
We assume that only packing classes $E$ are relevant that
satisfy the condition
\begin{equation} \label{spezialppc}
{\cal E}_{+,i} \subseteq E_{i}, \, i=1,\dots,d,
\end{equation}
where
${\cal E}_{+,1} \dots, {\cal E}_{+,d}$            
are given edge sets on $V$.

Consider a set of boxes 
$S \subseteq V$ that contains two boxes $b,c$, such that
$e=bc \in {\cal E}_{+,i}$.
With $e$, at least one of the edges of the clique
$\clique{S}$ must be contained in the
$i$th component graph of any relevant packing class.
This implies that $S$ can never occur as an independent set
of the $i$th component graph in condition (P2).
Therefore, $i$-feasibility of $S$ is not an issue,
so removal of $S$ from  ${\cal F}(V,w_{i})$ 
cannot delete any relevant packing classes.
This means that for the purpose of size modification, we only have to consider 
sets of boxes $S$ with cliques
$\clique{S}$ that are disjoint from ${\cal E}_{+,i}$.
The family of these sets is denoted by
$$
{\cal F}(V,w_{i},{\cal E}_{+,i}) := 
\left\{ S \in {\cal F}(V,w_{i}) | \clique{S} \cap {\cal E}_{+,i} = \emptyset \right\}.
$$
Therefore, the assumptions of Theorem~\ref{bilanz} can be weakened: 
\begin{satz} \label{bilanz2}
Let $(V,w)$ and $(V,w')$ be OPP-$d$ instances, 
let
${\cal E}_{+} = ({\cal E}_{+,1} \dots, {\cal E}_{+,d})$ be $d$-tuples of edge sets
on $V$, 
and let $E$ be a packing class of $(V,w)$ that satisfies (\ref{spezialppc}).
Suppose that for all $i \in \{1,\dots,d\}$ the following holds:   
\begin{equation} \label{bilcond2}
{\cal F}(V,w_{i},{\cal E}_{+,i}) \subseteq {\cal F}(V,w'_{i}).
\end{equation}
Then $E$ is a packing class for $(V,w')$.
\end{satz}
{\bf Proof.} 
Like in the proof for Theorem~\ref{bilanz} we only have to check condition (P2).
By (\ref{spezialppc}) any independent set $S$ of the $i^{th}$ 
component graph satisfies
$\clique{S} \cap  {\cal E}_{i} = \emptyset$.
Then the assumption (\ref{bilcond2}) guarantees that
(P2) remains valid when changing $w$ to $w'$.
\qed

\begin{definition}[Generalization of Definition \ref{konska}]
\label{konska2} 
Given the assumptions of Theorem \ref{bilanz2}.
Then we say that $w'$ is a {\em conservative scale} for $(V,w,{\cal E}_{+})$.
\end{definition}
Definitions
\ref{konska} and \ref{konska2} are compatible.
A conservative scale for $(V,w,(\emptyset,\dots,\emptyset ))$ is a
conservative scale for $(V,w)$.
Conversely, a conservative scale for $(V,w)$ 
is a a conservative scale for $(V,w,{\cal E}_{+})$
for any $d$-tuple of edge sets
${\cal E}_{+}$ of $V$.

Corollary \ref{notkor} can now be generalized:
\begin{coro} \label{notkor2}
Let $w'$ be a conservative scale for $(V,w,{\cal E}_{+})$. 
Then condition (\ref{volumencond}) from Corollary \ref{notkor}
is a sufficient condition for the existence of a
packing class for $(V,w)$ that satisfies condition (\ref{spezialppc}).
\end{coro}

So far we have used a selection of easily constructible conservative scales 
in order to apply the volume criterion.
Now we describe a method for improving conservative scales by
increasing the total box volume. Our approach uses the information
provided by (\ref{spezialppc}). The idea is to stretch an individual
box along a given coordinate as much as possible while preserving
a conservative scale.
\begin{lemma} \label{konskaknapsack}
Let $(V,w)$ be an OPP-$d$ instance and let ${\cal E}_{+}$ be a $d$-tuple of edge sets on $V$.
Let $b \in V$ be a box and let $i \in \{1,\dots,d\}$ be a coordinate direction.
Choose
\begin{equation} \label{ksks2}
\lambda \ge \max \left\{ w_{i}(S) \, | \,  S  \in {\cal F}(V,w_{i},{\cal E}_{+,i})
\, \mbox{ and } \, b \in S
  \right\} .
\end{equation}
Then the size function $w'$ given by
\begin{equation}
w'_{j}(c) := \left\{ \begin{array}{ll}
w_{j}(c) + (1 - \lambda) & \mbox{ for } (j,c) = (i,b) \\
w_{j}(c) & \mbox{ else }
\end{array} \right.
\end{equation}
is a conservative scale for $(V,w,{\cal E}_{+})$.       
\end{lemma}
{\bf Proof.} 
Condition (\ref{bilcond2}) needs to be checked only for coordinate
$i$ and for sets containing $b$.
Therefore consider a set $S  \in {\cal F}(V,w_{i},{\cal E}_{+,i})$ with $b \in S$. 
Then the choice of $\lambda$ assures that
$$
w'_{i}(S) = w_{i}(S) + (1 - \lambda) \le
w_{i}(S) + (1 - w_{i}(S)) = 1.\qed
$$
\nopagebreak

It suffices to consider an upper bound for the maximum in
(\ref{ksks2}) for these computations. For example,
the family
${\cal F}(V,w_{i},{\cal E}_{+,i})$ is contained in the the power set of 
$\{ b \} \cup  \{ c \in V \, | \, bc \in {\cal E}_{+,i} \}$.
Replacing the family by the power set yields an upper bound
that can be computed as the solution of a one-dimensional knapsack problem.

The following example illustrates the technical lemma and the underlying
terms.
\begin{exam} \label{konskabeispiel}
Consider the OPP-$2$ instance $(V,w,W)$ with
$$
V  :=  \left\{ 1',\dots,6' \right\}, \quad
W := (20,13),
$$
$$
\begin{array}{rclrcl}
w(1') = w(2') & := &  (8,7),&
w(3') & := &  (12,4),\\
w(4') = w(5') & :=  & (6,6),&
w(6') & := &  (8,3).\\
\end{array}
$$
Using 
Theorem~\ref{konskaknapsack}, we construct a conservative scale
that disproves the existence of a packing class with
the side constraints
$$E_{1} \supseteq {\cal E}_{+,1} := \{ 1'3', 4'5' \}$$.
 
We start by giving an intuitive idea of the situation.
Let direction $1$ be ``horizontal''
and let direction $2$ be ``vertical''.
For a packing class, consider the space right and left of box $1'$:
\begin{itemize}
\item Box $3'$ must be outside of this space, as it is adjacent to $1'$ in 
in ${\cal E}_{+,1}$; this implies a horizontal overlap with $1'$.
\item If $2'$ or $6'$ are packed alongside $1'$, horizontal
free space of size
$
W_{1}-w_{1}(1')-w_{1}(2')
= W_{1}-w_{1}(1')-w_{1}(6') =
20-8-8=4$ is left at the respective level, which does not fit any further boxes.
\item $4'$ and $5'$ would both fit next to $1'$.
Because of $4'5' \in {\cal E}_{+,1}$,
one of these boxes must lie above the other. 
This means that at most one of the boxes
$4'$ and $5'$ is packed alongside $1'$, leaving horizontal free space of size
$
W_{1}-w_{1}(1')-w_{1}(4')
= W_{1}-w_{1}(1')-w_{1}(5') =
20-8-6=6$.
\end{itemize}
At any level right and left of box $1'$, 
at least a total of $4$ units must remain free.
This implies an unpacked space of 
size at least $4 \cdot 7$.

These facts can be covered by the terminology of conservative scales: 

The maximal sets
from ${\cal F}(V,w_{1},{\cal E}_{+,1})$ that contain $1'$ are
$$
\{1',2'\}, 
\{1',4'\}, 
\{1',5'\}, 
\{1',6'\}.
$$

This implies
$$
\max \{ w_{1}(S) | S \in {\cal F}(V,w_{1},{\cal E}_{+,i}), \, 1' \in S \}  
\, = \, \max \{ 8+8, \, 8+8, \, 8+6, \, 8+6 \} \, = \, 16.
$$                 
By Theorem~\ref{konskaknapsack} we get from $w$ a conservative scale
for $(V,w,{\cal E}_{+})$by replacing $w_{1}(1')=8$ by $8+(20-16)=12$. 
With respect to this conservative scale, the total box volume is
$$
12 \cdot 7 +
8 \cdot 7 +
12 \cdot 4 +
6 \cdot 6 +
6 \cdot 6 +
8 \cdot 3 
 = 284 > 20 \cdot 13 = W_{1} \cdot W_{2}.
$$
By Theorem~\ref{notkor2}, no packing class with 
$E_{1} \supseteq {\cal E}_{+,1}$ can exist.
\qed
\end{exam}

\section{Conclusions}
\label{sec:conc}
We have described a new framework for obtaining lower bounds
for higher-dimensional packing problems. We have shown
that all known classes of lower bounds for these problems
can easily be formulated in and improved by this framework.
In this paper we have given three
particular classes of dual feasible functions.
Furthermore, any additional set of dual feasible functions
for one-dimensional packing (described in detail in
our paper~\cite{FESE97}) can be used immediately for
constructing new lower bounds for higher-dimensional
packing problems, by using convex combinations and
compositions. 

When considering the performance of bounds resulting
from dual feasible functions, one should also realize 
the limitations: As any such bound considers only one item
at a time (which is why the bounds have linear complexity
after sorting), complications resulting from the more involved
{\em combination} of items cannot explicitly be recognized. Again, see
\cite{FESE97} for a discussion in the context of one-dimensional
bin packing. Thus, a possible way to 
stronger bounds may be to consider logical implications
of several packed items at a time.

We omit a separate computational study on the performance
of our new bounds. Instead, we demonstrate their 
usefulness by applying them to actually solve
higher-dimensional packing problems to optimality.
Combining our new classes of bounds with our characterization
of feasible packings (described in \cite{pack1,fs-cchdop-04}),
we get a powerful method that can solve instances
of previously unmanageable size. All technical details
are described in our paper \cite{pack3}.

\section*{Acknowledgment}
We thank an anonymous referee for comments that helped in
preparing the final version of this paper.

\clearpage
\bibliographystyle{plain}
\bibliography{packing}

\end{document}